\documentstyle[prd,aps]{revtex}
\begin{document}
\input epsf
\draft
\twocolumn[\hsize\textwidth\columnwidth\hsize\csname
@twocolumnfalse\endcsname
\preprint{SU-ITP-98-15, gr-qc/9803068}
\title{Quantum Creation of  a  Universe with $\Omega \not =1$:
Singular and
Non-Singular Instantons}
\author{Raphael Bousso and Andrei Linde}
\address{Department of Physics, Stanford University, Stanford, CA
94305, USA}
\date{ 19 March, 1998}
\maketitle
\begin{abstract}
  We propose two new classes of instantons which describe the
  tunneling and/or quantum creation of closed and open universes.  The
  instantons leading to an open universe can be considered as
  generalizations of the Coleman-De-Luccia solution. They are
  non-singular, unlike the instantons recently studied by Hawking and
  Turok,  whose prescription has the problem that the singularity
  is located on the hypersurface connecting to the Lorentzian region,
  which makes it difficult to remove. We argue that such singularities
  are harmless if they are located purely in the Euclidean region. We
  thus obtain new singular instantons leading to a closed universe;
  unlike the usual regular instantons used for this purpose, they do
  not require complex initial conditions.  The singularity gives a
  boundary contribution to the action which is small for the
  instantons leading to sufficient inflation, but changes the sign of
  the action for small $\phi$ corresponding to short periods of
  inflation.

\end{abstract}
\pacs{PACS: 98.80.Cq  \hskip 3.6cm SU-ITP-98-15
\hskip 3.6cm  gr-qc/9803068}
\vskip2pc]

\section {Introduction}

It is well known that most of the models of inflationary cosmology
predict $\Omega = 1 \pm 10^{-4}$. It is possible to have inflation
with $\Omega \not = 1$, but it is rather difficult. The basic idea is
to solve the homogeneity and isotropy problem not by the long stage of
inflation, but by quantum tunneling to a state describing an open or
closed universe. Then the universe will be homogeneous if the
probability of tunneling is sufficiently strongly suppressed.  In this
scenario an infinite number of open universes can be created in one of
two ways. One may consider a purely classical evolution of an
inflationary universe in the false vacuum and a subsequent creation of
inflating open universes by tunneling to the true
vacuum~\cite{Gott,BGT,Open}.  Alternatively, one may consider the
quantum creation from nothing of a closed inflationary universe, which
later decays into   an infinite number of open universes by the
process described above \cite{Erice,ALOpen}. In all such models, it is
necessary to assume a potential with a false vacuum.

Recently, Hawking and Turok claimed that open universes can be
obtained without an intermediate stage involving false vacua.  They
described a process in which an open universe is created from nothing
in the chaotic inflation scenario with a generic effective
potential~\cite{HT}. They used the standard deformed-four-sphere
Euclidean solution, in which the inflaton field is constant on the
lines of constant latitude. This solution generically has a
singularity on one of the poles. Usually it is cut along the equator;
the singular hemisphere is discarded, and the regular one is
analytically continued to yield a closed Lorentzian universe.

Instead, Hawking and Turok cut the Euclidean solution through the
poles, thus including the singularity on the hypersurface through
which the Euclidean and Lorentzian sectors are joined. The
hypersurfaces of constant inflaton field will then form infinite open
spacelike sections in a part of the resulting Lorentzian universe.

This approach suffers from two problems. The first problem is that
they obtain $\Omega = 10^{-2}$ for the ratio of the present density to
the critical density.  This contradicts observational data by almost
two orders of magnitude, and even this result was obtained only after
invoking the anthropic principle, without which one would get $\Omega
= 0$. The prediction comes from the probability measure associated
with the Hartle-Hawking wave function. In \cite{ALOpen} it was argued
that, according to~\cite{Creation,book}, this wave function should not
be used for the description of the creation of the universe; and it
was shown that the use of the tunneling proposal would typically lead
to $\Omega \approx 1$.  It thus appears that with either choice of the
wave function of the universe we are not currently in a position to
obtain a realistic value of $\Omega=0.3$ unless customized potentials
are employed.

The second problem is associated with the presence of a singularity on
the nucleation surface. Vilenkin~\cite{VIL} argued that instantons of
the Hawking-Turok type lead to vacuum instability and should therefore
be excluded from the path integral. In \cite{ALOpen} it was shown that
not every instanton is permitted even if it is non-singular. On the
other hand, it was suggested that singular instantons are not
necessarily forbidden, but one should be extremely careful about the
analytical continuation involving singularities which was used in
\cite{HT}.  For other problems associated with this issue see also
\cite{Unruh}.

In this paper, we will perform a more detailed investigation of these
issues. We will suggest two ways of avoiding the problems associated
with the Hawking-Turok singularity. First, we will consider potentials
with a local maximum, for which there are non-singular solutions. They
include the Coleman-De-Luccia instantons as well as some new, related
solutions which we found. We will discuss the structure and
application of these solutions in Sec.~\ref{sec-nonsing}. They
describe the nucleation of open universes, and allow the correct
prediction of $\Omega$ for suitable potentials. Of course, this means
that the generality claimed by Hawking and Turok is lost, but as we
pointed out above, generic inflaton potentials do not seem very
promising in any case when one tries to predict universes which are
both non-flat and non-empty.

We will allow generic potentials in Sec.~\ref{sec-sing}, where we will
use variants of the deformed-four-sphere instanton to nucleate closed
universes. We cut them along the equator and discard the {\em regular}
hemisphere. The Hawking-Turok singularity will be present in this
case. It will not, however, lie on the nucleation hypersurface.
Therefore it can be ``surgically removed,'' or viewed as a small
region of Planckian density. We calculate the boundary contribution to
the action and show that it is small in all cases where sufficient
inflation ensues. We discuss a possible interpretation of these
solutions as the birth of a closed inflationary universe by tunneling
from space-time foam. Finally, we construct instantons which are
symmetric about the equator and contain two singularities. They allow
the construction of nucleation paths on which all variables are
everywhere real.

\section{Non-Singular instantons} \label{sec-nonsing}

\begin{figure}[Fig0]
 \hskip 1.5cm
\leavevmode\epsfysize=4cm \epsfbox{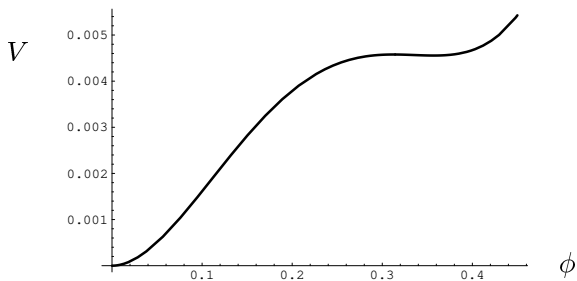}

\

\caption[Fig1]{\label{potential} Effective potential
  $V(\phi) = {m^2\over 2}(\phi^2(\phi- v)^2 +B \phi^4)$ for $m^2 = 2$,
  $B=0.12$ and $v = 0.5$.  It has a shallow minimum at   $\phi_0 =
  0.357$ and a local maximum at $\phi_1=0.312$. All quantities in this
  figure are in units of $M_{\rm p}/\sqrt{8\pi}$.}
\end{figure}

Suppose we have an effective potential $V(\phi)$ with a local minimum
at $\phi_1$, and a global minimum at $\phi=0$, where $V=0$ (see Fig.
\ref{potential}).  In an $O(4)$-invariant Euclidean spacetime with the
metric
\begin{equation}\label{metric}
ds^2 =d\tau^2 +a^2(\tau)(d \psi^2+ \sin^2 \psi \, d \Omega_2^2) \ ,
\end{equation}
the scalar field $\phi$ and the three-sphere radius $a$ obey the
equations of motion
\begin{equation}\label{equations}
\phi''+3{a'\over a}\phi'=V_{,\phi},~~~~~ a''= -{8\pi G\over 3} a (
\phi'^2 +V) \ ,
 \end{equation}
where primes denote derivatives with respect to $\tau$.

These equations have several non-singular solutions, the simplest of
which are the $O(5)$ invariant four-spheres one obtains when the field
$\phi$ sits at one of the extrema of its potential. In this case the
first of the two equations above is trivially satisfied, and $a(\tau)
= H^{-1} \sin H\tau$. Here $H^2 = {8\pi V\over 3 M_{\rm p}^2}$.   
Using the solution for which $\phi=\phi_1$, Hawking and Moss \cite{HM}
found the rate at which the field $\phi$ in a single Hubble volume
tunnels to the top of the potential, from which it can roll down
towards the true vacuum.  For a recent discussion of this instanton
and its interpretation see \cite{ALOpen}. The main other use of these
trivial instantons is to find the action of the false vacuum
background solution, which must be subtracted from the bounce action
to obtain a tunneling rate.

We shall consider potentials for which $V_{,\phi\phi} \gg H^2$ in the
region where the tunneling occurs. In this case, tunneling out of the
false vacuum does not occur primarily on the scale of an entire Hubble
volume via the Hawking-Moss instanton. Instead the transition will
proceed via more complicated Euclidean solutions with varying field
$\phi$. These include the  Coleman-De-Luccia instanton, and related
instantons which we found.

\subsection{Bubble instantons}

\begin{figure}[Fig1]
 \hskip 1.5cm
\leavevmode\epsfysize=9.5cm \epsfbox{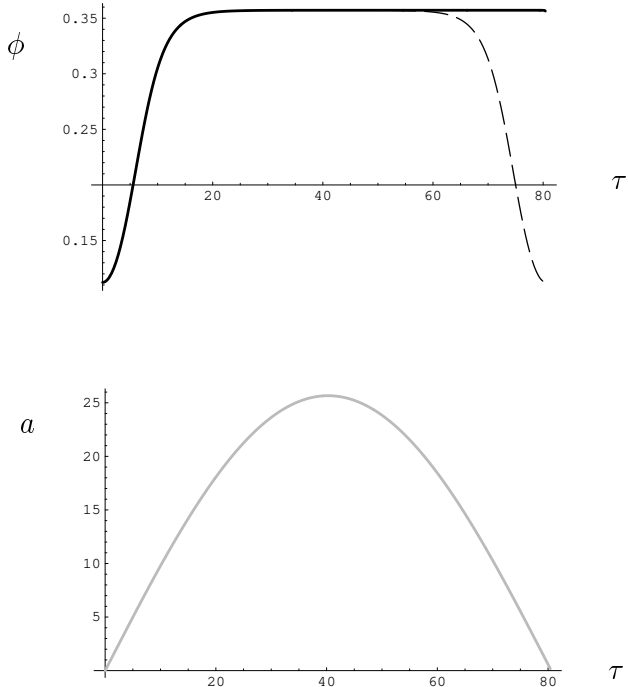}

\

\caption[Fig1]{\label{Colem} The upper panel shows the behavior of
  the scalar field $\phi$ for examples of the Coleman-De-Luccia
  ``bubble'' instanton (solid line) and the new ``double-bubble''
  instanton which we have found (dashed line).  For both instantons,
  the field is in the domain of the true vacuum at small $\tau$,
  forming a bubble.  For the bubble instanton, the field is closest to
  the false vacuum at the pole opposite the bubble. For the
  double-bubble instanton, this happens on the equator, at the moment
  of the maximal expansion. The behavior of the three-sphere radius
  $a(\tau)$ shown in the lower panel is very similar for both
  instantons, though it is not identical.}

\end{figure}

A Euclidean solution which describes the creation of an open universe
was first found by Coleman and De Luccia in 1980 \cite{CL}.  It is
given by a slightly distorted de~Sitter four-sphere of radius
$H^{-1}(\phi_0)$. Typically, the field $\phi$ is very close to the
false vacuum, $\phi_0$, throughout the four-sphere except in a small
region (whose center we may choose to lie at $\tau=0$), in which it
lies on the `true vacuum' side of the maximum of $V$. The behavior of
the field and scale factor for the potential in Fig.~\ref{potential}
is shown in Fig.~\ref{Colem}. The scale factor vanishes at the points
$\tau=0$ and $\tau=\tau_{\rm f} \approx \pi/H$, which we will call the
North and South pole of the four-sphere. In order to get a
singularity-free solution, one must have $\phi' = 0$ and $a'=\pm 1$ on
the poles.

\begin{figure}[Fig1]
 \hskip 1.5cm
\leavevmode\epsfysize=5cm \epsfbox{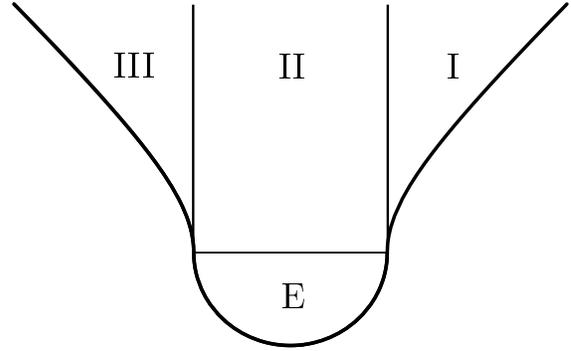}

\

\caption[Fig1]{\label{fig-regions}
  The Lorentzian de~Sitter-like spacetime obtained from the analytic
  continuation of Coleman-De-Luccia instantons contains three regions.
  In Regions I and III the hypersurfaces of constant field $\phi$ form
  open spacelike sections. Region II is a shell separating the two
  bubbles.}

\end{figure}

This solution can be cut in half along the line $\psi=\pi/2$, which
removes half of each three-sphere. Then one can continue analytically
to a Lorentzian spacetime~\cite{GutWei83,HT} with the time variable
$\sigma$, given by $\psi=\pi/2+i\sigma$. This gives region II of the
Lorentzian universe (see Fig.~\ref{fig-regions}):
\begin{equation}
ds^2 = -a^2(\tau)\ d\sigma^2 + d\tau^2 + a^2(\tau) \cosh^2 \sigma\
d\Omega_2^2;
\end{equation}
the field $\phi$ will still depend on $\tau$ in the same way as
before, and will be independent of $\sigma$.  This describes a shell
of width $H^{-1}$, which is mostly near the false vacuum and expands
exponentially. The shell separates two bubbles, regions I and III, in
which the universe looks open.

Region I is obtained by taking $\sigma = i\pi/2 + \chi$ and $\tau =
it$, giving the metric
\begin{equation}
ds^2 = -dt^2 + \alpha^2(t) \left( d\chi^2 + \sinh^2 \chi d\Omega_2^2
\right),
\end{equation}
where $ \alpha(t) = -i\, a[\tau(t)]$. Its spacelike sections (defined
by the hypersurfaces of constant inflaton field) are open. Thus,
region I looks from the inside like an infinite open universe, which
inflates while the field $\phi$ slowly rolls down to the true vacuum.
The evolution will then undergo a transition to a radiation or
matter-dominated open Friedman-Robertson-Walker universe.

In region III, which is obtained by choosing $\sigma = i\pi/2 + \chi$
and $\tau = \tau_{\rm f} + it$, the field $\phi$ rolls to the local
minimum at $\phi_0$, and one gets indefinite open inflation in the
false vacuum.

The analytic continuations we have given support the interpretation of
such solutions as the spontaneous nucleation of a bubble of true
vacuum on the background of de~Sitter space expanding in the false
vacuum. For this reason we will call them `bubble instantons'.  The
nucleation rate is given by
\begin{equation}
\Gamma = e^{-\Delta S},
\end{equation}
where $\Delta S$ is the difference between the action of the full
Euclidean bubble solution, and the action of a Euclidean solution
describing the background spacetime.  Except for near-Planckian
potentials, both actions will be large and negative (about   $-2.6
\times 10^4$ in our example).  The background solution is given by an
exact Euclidean four-sphere on which the field $\phi$ is constant and
equal to $\phi_0$, the false vacuum. Its action will be $-{3 M_{\rm
    p}^4 \over 8 V(\phi_0)}$.  Subtracting   this from the action of
the bubble solution, one obtains a positive $\Delta S$ ($\approx 4.9$
in our example). This means that bubble formation by tunneling is
suppressed, as it should be.

One usually requires instanton solutions to interpolate between the
initial and final spacelike sections (in this case, a section of pure
de~Sitter space in the false vacuum and a similar section containing a
bubble of true vacuum). The above description, which seems to use two
disjoint instantons, is actually consistent with this formal
requirement, since the instantons may be connected by virtual
domain walls after small (Planck size) four-balls are removed. This
will cause the background instanton to contribute to the total action
with a negative sign. If one connects the background instanton to the
region of the bubble instanton where $\phi$ is closest to its false
vacuum, the discontinuity in $\phi$ will be small, so the volume
contributions of the removed regions cancel almost exactly. Requiring
continuous instantons, therefore, does not change the pair creation
rate significantly~\cite{BC}.

Cosmological instantons have frequently been interpreted to describe
the creation of a universe from nothing, i.e.\ without a pre-existing
background. This case is considerably less well-defined than the
quantum nucleation of structures on a given background solution. In
particular, the sign with which the large, negative action enters the
exponent in the path integral is subject to
debate~\cite{HT,ALOpen,HTnew}. Leaving such questions aside for now,
we will take the position that isolated cosmological instantons are
indeed related to universe creation, independently of the formalism
used to assign probabilities to such processes.

\subsection{Double-bubble instantons}

We have found a new instanton in which there are two bubbles, one on
each pole. In this solution, $\phi$ is in the domain of the true
vacuum in small regions near the poles, and near the false vacuum
elsewhere; this can be seen from the dashed line in Fig.~\ref{Colem}.
The geometry is still approximately a four-sphere. As before, $\phi'$
vanishes on the poles; but now it also vanishes on the equator, at
$\tau=\tau_{\rm max}$. The Northern and Southern hemispheres are
exactly symmetric.

Not surprisingly, the action of the double-bubble solution, after the
background subtraction described above, is approximately twice that of
the bubble (Coleman-De-Luccia) instanton.  For the instanton shown in
Fig.~\ref{Colem} one has $\Delta S_2 \approx 9.8$.

The analytic continuations will be the same as before, with a
different result. Region II will be mostly in the domain of the false
vacuum. Region I and III will be identical, each containing an open
inflating universe in which the field rolls down to the true vacuum.
Globally, therefore, we obtain two bubbles of true vacuum separated by
a shell which inflates in the false vacuum.

This solution can be interpreted as the spontaneous pair-creation of
bubbles of open inflation on the background of false vacuum inflation.
Alternatively, one may view it as the creation from nothing of two
open inflating universes separated by a metastable shell.

\subsection{Anti-double-bubble instantons}

\begin{figure}[Fig0111]
 \hskip 1.5cm
\leavevmode\epsfysize=4.5cm \epsfbox{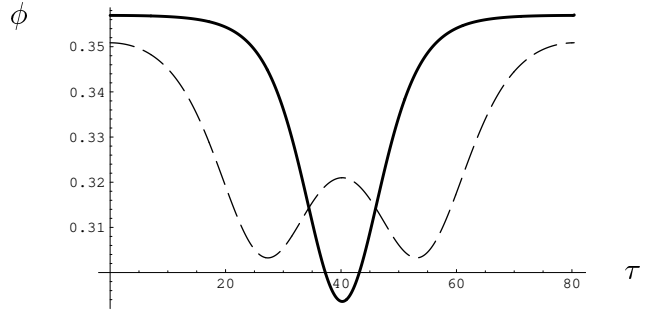}

\

\caption[Fig1]{\label{loop} Two examples of ``anti-double-bubble''
  instantons, in which the field is in the false vacuum domain near
  the poles, and reaches into the domain of the true vacuum on a shell
  near the equator. It can be cut through the poles to describe shell
  nucleation, or across the equator, describing the tunneling to
  true-vacuum inflation in a closed universe.}

\end{figure}

In addition we have found another family of instantons, two examples
of which are shown in Fig.~\ref{loop}. In these instantons, the field
is in the domain of the false vacuum in two regions surrounding the
poles. They are separated by a thin shell at the equator, where the
field is in the true vacuum domain. These instantons have a much
greater action difference to the background instanton, since the
true-vacuum region is significantly larger than in the previous two
cases. In particular, $\Delta S = 93.6$ for the instanton shown by the
solid line in Fig.~\ref{loop}, and $\Delta S = 124.7$ for the
instanton shown by the dashed line.

\subsubsection{Open cut}

With the analytic continuation used for the previous two instantons,
regions I and III will become open inflationary universes in which the
field rolls down to the false vacuum. They are separated by the region
II, which contains a shell on which $\phi$ is in the domain of the
true vacuum.

Therefore we may interpret this solution as the nucleation of a shell
of true vacuum on a false vacuum inflationary background, or
alternatively, as the creation of such a universe from nothing.
Because of the larger action difference, spontaneous shell creation
will be quite suppressed compared to bubble formation.

\subsubsection{Closed cut}

A more intriguing application of this instanton can be found by
choosing a different analytic continuation. Instead of cutting at
$\psi=\pi/2$, we may choose to leave the three-spheres intact, and cut
across the equator. Lorentzian time will be defined by $\tau=\tau_{\rm
  max} + iT$, and we obtain a metric with closed spacelike sections:
\begin{equation}
ds^2 = -dT^2 + a^2(T) d\Omega_3^2.
\label{eq-closed}
\end{equation}
The inflaton field is in the domain of the true vacuum on the
nucleation surface (the equator), so it will start rolling down
towards the absolute minimum. During this time, the spacelike
three-spheres grow exponentially:
\begin{equation}
a(T) \approx H^{-1}(T) \cosh \int H(T)\, dT.
\label{eq-cosh}
\end{equation}
Thus we obtain a closed inflationary universe in which the scalar
field rolls towards the true vacuum.

One could interpret this instanton as describing the creation of such
a universe from nothing. But this would just add an alternative to the
usual instantons on which the field is entirely in the domain of the
true vacuum. A much more interesting interpretation is the one
associated with a pre-existing background of false vacuum inflation.
In this case, the anti-double-bubble instanton is seen to describe the
spontaneous tunneling to the true vacuum in an entire Hubble-volume of
de~Sitter space. Unlike the Coleman-De-Luccia bubbles, these regions
will not contain an open universe. In the example we are considering,
for which Hawking-Moss tunneling is not possible, this shows that one
can nevertheless nucleate true vacuum bubbles containing a closed
universe.

\section{Singular Instantons} \label{sec-sing}

\subsection{Standard instantons}

\begin{figure}[Fig0111]
 \hskip 1.5cm
\leavevmode\epsfysize=6.5cm \epsfbox{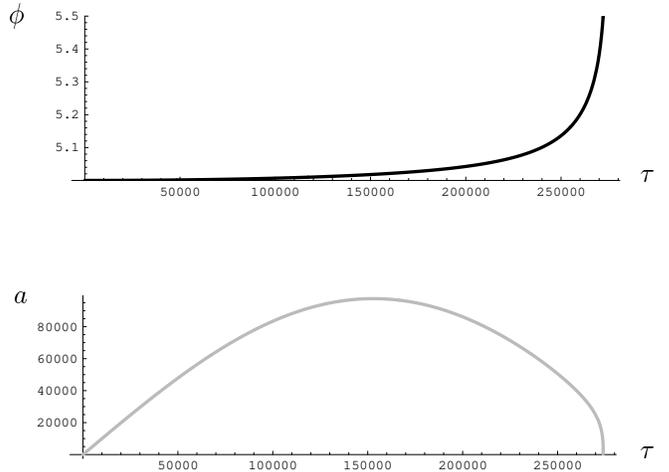}

\

\caption[Fig1]{\label{fig-trouble}
  The standard solution. The field $\phi$ as well as the curvature are
  singular at the South pole.  Note that $\phi'$ does not vanish at
  the equator, so one should complexify the scalar field in order to
  make the analytical continuation to a closed universe.  The
  analytic continuation to an open universe suggested by Hawking and
  Turok~\cite{HT} involves the singularity on the nucleation surface,
  which makes it rather problematic.}

\end{figure}

We now assume a generic effective potential of chaotic inflation, with
a minimum, $V=0$, at $\phi=0$, and no other stationary points. The
standard Euclidean solution used in quantum cosmology is obtained by
requiring regularity at the North pole, at $\tau=0$. This means one
must take $a'=1$ and $\phi'=0$ there. A Euclidean solution with these
initial conditions in shown in Fig.~\ref{fig-trouble} for a massive
scalar field.

On most of the manifold, the solution will be almost a
four-sphere, with $\phi$ increasing very slowly:
\begin{equation}
a(\tau) = H^{-1} \sin H \tau,~~~~~~
\phi(\tau) = \phi_{\rm N},
\end{equation}
where $\phi_{\rm N}$ denotes the value of the inflaton field on the
regular pole.  A first approximation to the Euclidean action of the
standard instanton will therefore be given by the volume term for a
four-sphere of radius $H^{-1}(\phi_{\rm N})$, $S \sim - {3 M_{\rm p}^4
  \over 8V(\phi_{\rm N})}$.  This will be a good approximation for
large values of $\phi_{\rm N}$ leading to long periods of inflation.

Near the South pole, at $\tau=\tau_{\rm f}$, the anti-damping term
$\sim \phi' b'/b$ starts to dominate the equation of motion for
$\phi$. The field diverges logarithmically, and the potential terms
can be neglected. Approximate solutions are given by \cite{HT,VIL}
\begin{eqnarray}
a(\tau) & = & A (\tau_{\rm f}-\tau)^{1/3},
\label{eq-nearsing} \\
\phi(\tau) & = & - \frac{1}{\sqrt{12\pi}}
 \ln (\tau_{\rm f}-\tau) + \phi_{\rm m},
\end{eqnarray}
where $A$ and   $\phi_{\rm m}$ are constants.

There are many questions associated with  this singularity. First
of all, even though the divergence of the scalar field is only
logarithmic, the energy density and curvature diverge according to a
power-law. If the singularity is part of the nucleation geometry,
i.e.\ if it is included in the Euclidean instanton and its Lorentzian
analytic continuation, the corresponding method can no longer be
called ``the no-boundary proposal.''

The boundary at $\tau_{\rm f}$ contributes a
Gibbons-Hawking \cite{GibHaw77b} term to the action:  
\begin{equation}
S_{ \tau_{\rm f} } = -\frac{M_{\rm p}^2}{8\pi} \left.
  \int d^3\!x\, h^{1/2}
  K \right|_{\tau=\tau_{\rm f}},
\end{equation}
where $h$ is the determinant of the three-metric $h_{ij}$, and $K$ is
the trace of $K_{ij}$, the second fundamental form.  For an
$O(4)$-invariant metric, we find $K = 3 a'/a$. This yields  
\begin{equation}
S_{ \tau_{\rm f} } = \left.
- \frac{\pi M_{\rm p}^2}{4}
 \frac{d(a^3)}{d\tau}\right|_{\tau=\tau_{\rm f}}  .
\end{equation}
By Eq.~(\ref{eq-nearsing}), $a(\tau)$ goes like $(\tau_{\rm
  f}-\tau)^{1/3}$ near the singularity. Therefore, the boundary term
will be positive and finite.\footnote{This result was also
  independently found by Vilenkin \cite{VIL}.} Note that for any power
other than $1/3$, the boundary term would either vanish, or diverge.

The contribution of this boundary term to the total action is
relatively small if one considers the creation of an inflationary
universe. For example, numerical investigation shows that in the
theory ${m^2\phi^2/ 2}$ this correction, as compared to the action
$-{3 M_{\rm p}^4 \over 8 V(\phi_{\rm N})}$, is suppressed by  a
factor $O({M_{\rm p}\over \phi_{\rm N}})$ for $\phi_{\rm N} \gg M_{\rm
  p}$:  
\begin{equation}
S_{\rm std}
\approx
-{3 M_{\rm p}^4\over 8 V(\phi_{\rm N})}\left(1 - {M_{\rm p} \over
2\phi_{\rm N}}\right).
\end{equation}
There is also a correction to the volume term because $\phi$ is not
exactly constant anywhere; we will not discuss this correction here.

As is obvious from this result, at small $\phi_{\rm N} $ the total
action including the boundary term may become positive. One can
confirm numerically that this is indeed the case. This is a rather
unexpected conclusion as it indicates that the absolute value of the
action reaches its maximum not at $\phi_{\rm N} = 0$ but at $\phi_{\rm
  N} \sim M_{\rm p}$.

\subsubsection{Closed cut} \label{sec-closedcut1}

This Euclidean solution can be cut in different ways to obtain an
instanton that allows an analytic continuation to a Lorentzian
spacetime. The standard method is to cut along the equator, at $\tau=
\tau_{\rm max}$ and to discard the Southern hemisphere, which contains
the singularity. With $\tau = \tau_{\rm max} + iT$, one can join the
regular hemisphere across the equator to Lorentzian de~Sitter space
with the metric given by Eqs.~(\ref{eq-closed}) and
(\ref{eq-cosh}). The surfaces of constant inflaton field will be
closed spacelike slices in the Lorentzian sector. Therefore this cut
corresponds to closed inflation.

In general, $\phi'$ will be small but non-zero on the equator. This
means that it will acquire an imaginary part in the Lorentzian sector.
But at late Lorentzian times, when measurements are made, one must
demand that all variables be exactly real. It is therefore necessary
to compensate by adding a small imaginary part to the initial value of
$\phi$~\cite{Lyo92,BouHaw95}
\begin{equation}
\phi_{\rm N} = \phi_{\rm N}^{\rm Re}
  - i \frac{M_{\rm p}^2}{8\phi_{\rm N}^{\rm Re}}
\end{equation}
where the superscript ${\rm Re}$ denotes the real part of a quantity.

For nucleation geometries which are everywhere real, the real part of
the Euclidean action comes entirely from the Euclidean sector, since
the Lorentzian sector contributes only a purely imaginary part. But in
the current case, the Lorentzian sector will not be purely real for a
time of order $H^{-1}$ after the nucleation hypersurface. It will
therefore give a further correction to the real part of the Euclidean
action. It can be easily checked both analytically and numerically
that the ratio of this term to the the total action is of order
$\phi_{\rm N}^{-2}$. This corrects claims in~\cite{Lyo92} that the
correction is of the same order as the total action.

\subsubsection{Open cut}

Hawking and Turok~\cite{HT} have suggested to use the same analytic
continuation for the standard solution that had been traditionally
used for the Coleman-De-Luccia solution, and which we gave explicitly
in Sec.~\ref{sec-nonsing}. This involves cutting the Euclidean space
through the poles, thus including half of the singularity on the
nucleation surface. The resulting Lorentzian solution contains regions
I and II of Fig.~\ref{fig-regions}, while region III is cut off by the
singularity. In region I the hypersurfaces of constant $\phi$ trace
out infinite open spacelike sections.

  The interpretation of this instanton and the analytical continuation
proposed in \cite{HT} may be rather problematic. It was argued in
\cite{ALOpen} that even if instantons are nonsingular, they do not
always describe tunneling. But in this case one has additional
problems associated with cutting the singularity in half and
performing the analytical continuation there \cite{ALOpen}.

 Vilenkin~\cite{VIL} has argued that the admission of such
nucleation geometries, in which there is a singularity on the
hypersurface of vanishing second fundamental form, leads to problems
of vacuum instability. Such instantons exist for a Minkowski space
background, where they may cause an almost unsuppressed nucleation of
singular bubbles spreading out nearly at the speed of light. Clearly,
this is physically unacceptable. It poses a problem for the
prescription suggested by Hawking and Turok unless one finds sound
arguments why such instantons should be admitted for inflationary
universes, but not for flat space.

According to~\cite{VIL}, singular instantons may not be allowed at all
because they are not true (nonsingular) solutions of   the equations
of motion, which would correspond to an extremum of the action.
However, the singularity is not really a part of the manifold.
Moreover, one can sometimes cut it out, and consider a configuration
which is nonsingular but coincides with a singular instanton
everywhere except for a small vicinity of the singularity. If the
action of the instanton converges, then for a sufficiently small size
of the patch replacing the singular region, the action will differ
from the instanton action by less than $\Delta S = 1$. Such ``almost
solutions'' are perfectly admissible and play the same role in the
functional integral as the true solutions, see e.g.~\cite{Tunn,BC}.
However, if one makes an analytical continuation through the
singularity, one cannot easily remove it by the method described
above, and then it may pose a real problem.

A possible way out of this problem would be to use non-singular
instantons of the type we described in Sec.~\ref{sec-nonsing}. They
require special potentials with a false vacuum. This has the advantage
that one can, in principle, obtain any given value $\Omega <1$. The
price we pay is some loss of generality\footnote{There is a second,
  more brute-force way of eliminating the singularity: One may take a
  spherical region around the regular pole and join it to its mirror
  image across a domain wall of positive energy density. This method,
  which will be described in a separate publication~\cite{BouCha98},
  does not require false vacua, but assumes the presence of fields
  supporting the topological defect.}.

An interesting result which appears after the boundary term is taken
into account is that the action at very small $\phi$ changes its sign
and becomes positive. This means that the maximal absolute value of
the action is reached not at the point when $V(\phi) = 0$, but at some
other point, where inflation is still possible. It would be very
interesting if this point were at a sufficiently large value of
$\phi$, which would provide a realistic value of $\Omega$ within the
Hartle-Hawking approach without any use of the anthropic principle.
Unfortunately, however, our numerical investigation of this question
shows that in all realistic models with potentials $\sim \phi^n$, the
absolute value of the action is maximal at $\phi \lesssim M_{\rm p}$, which
does not lead to long inflation, and which, consequently, yields an
exponentially small value of $\Omega$.  For example, in the simplest theory ${m^2\phi^2\over 2}$ the absolute value of the action is maximal at $\phi \sim 0.6 M_{\rm p}$, which practically does not lead to any inflation whatsoever.

\subsubsection{Closed cut revisited} \label{sec-closedcut2}

We will now turn to a different possibility. We wish to study the
consequences of allowing the singularity to be part of the nucleation
geometry but not of the hypersurface joining the Euclidean and
Lorentzian section. The simplest such instanton is obtained by cutting
the standard solution, once again, across the equator, but discarding
the {\em regular} hemisphere, and keeping the {\em singular}
hemisphere.

The instanton thus corresponds to the interval $ \tau_{\rm max} \leq
\tau \leq \tau_{\rm f} $.  The Lorentzian section is obtained by
taking $\tau = \tau_{\rm max} + iT$, so it will be the same as that in
Sec.~\ref{sec-closedcut1}: a closed inflationary universe.

The Euclidean region looks mostly like a four-sphere of radius $\sim
H^{-1}(\phi_{\rm N})$. But there will be a region near the singularity,
where the curvature and energy density diverge. We may impose a
Planck-scale cut-off here, and think of the singularity and its
vicinity as a small Planckian region immersed in the large
four-sphere. The presence of this region on the South pole is the
crucial difference between this nucleation geometry and the one
studied in Sec.~\ref{sec-closedcut1}.

How should we interpret this difference? The regular instanton, viewed
in isolation, has often been interpreted as representing the creation
of the universe from nothing. This was motivated by its self-contained
nature; one might think of `nothing' as the vanishing of the scale
factor $a$, which occurs on the regular North pole. In contrast, the
interpretation of the singular hemisphere actually seems less vague.
We can think of this instanton as an interpolation between a Planckian
regime, and a large closed inflating universe. Therefore we propose
that it describes the spontaneous ignition of inflation from a bubble
of spacetime foam. In fact, this agrees with the interpretation of
creation from `nothing' proposed in \cite{Creation}. We are speaking
about a state where the classical part of metric strongly fluctuates,
so that one cannot measure distance using any measuring devices. This
is what may happen at the Planckian epoch. But at the Planck time one
would expect all physical fields to take large and strongly
fluctuating values, rather than a definite value corresponding to the
North pole of the usual nonsingular de Sitter instanton at $\tau = 0$.
In this sense the use of singular instantons seems quite appropriate
for the description of quantum creation of a closed universe from
space-time foam.

The main difference of this use of the singular instanton to that
proposed by Hawking and Turok is that the singularity in our case does
not reach into the Lorentzian sector. It is limited to a tiny region
in the Euclidean regime, where it can easily be smoothed out, or
removed. Since  neither the boundary term nor the volume term have
divergences near the singularity, the action will be finite. For large
$\phi_{\rm N}$ it will not differ noticeably from the action of the
other hemisphere, or the action of the Hawking-Turok instanton.

\subsection{Gondola instantons}

In the previous subsection we argued that weak localized singularities
inside the Euclidean sector of a tunneling geometry can be interpreted
as interpolations to space-time foam and can thus be quite useful.
Once this point of view is adopted, however, it is easy to see that
the standard Euclidean solution is only a special case in a
one-parameter family of solutions. Generically, these solutions will
have singularities on both poles.

\begin{figure}[Fig0111]
 \hskip 1.5cm
\leavevmode\epsfysize=7.5cm \epsfbox{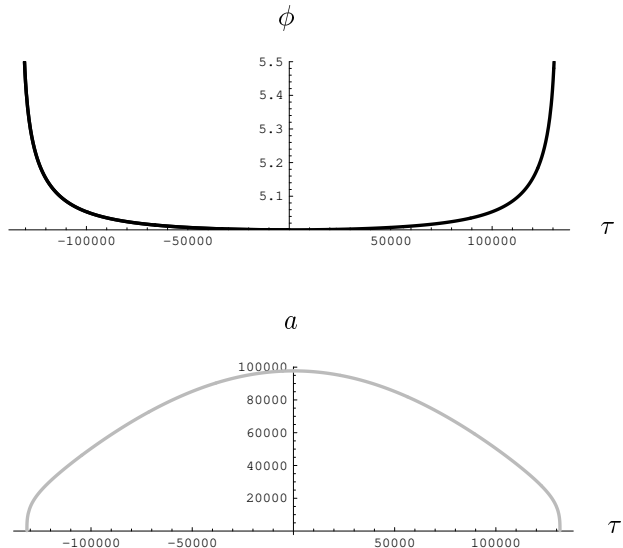}

\

\caption[Fig1]{\label{fig-gondola} Gondola solution.
  The field $\phi$ as well as the curvature are singular at the South
  and North poles. Unlike the standard solution, this one is symmetric
  about the equator, where all derivatives vanish.}

\end{figure}

We will now focus on a particular member of this family that is
exactly symmetric about the equator, shown in Fig.~\ref{fig-gondola}.
It can be constructed by specifying very simple boundary conditions on
the equator: One is free to choose the initial value of the field,
$\phi=\phi_{\rm E}$; the derivatives of all fields and metric
components are set to zero.  There will thus be identical
singularities on the North and South pole. We will call this the
`gondola' solution.

If we cut this solution along the equator, we obtain two identical
hemispheres, each containing a small Planckian region at its pole. As
we discussed above, we may consider this region to interpolate to
spacetime foam.  Performing the usual analytic continuation, $\tau =
\tau_{\rm max} + iT$, we obtain, once more, a closed Lorentzian
universe.  But the gondola instantons have the great advantage that
the second fundamental form, and all field derivatives, vanish on the
nucleation hypersurface by construction. This means that all variables
will be perfectly real in the entire Euclidean and Lorentzian sectors.
There is no need for introducing complex initial conditions in this
case.

The gondola solution has two boundaries which contribute terms to the
action. For comparable values of $\phi_{\rm E}$ on the equator, we
found numerically that these terms add up almost exactly to the
contribution of the single boundary term in the standard solution. The
instanton given by half of the gondola solution will contain only one
Planckian boundary. Therefore, compared to the singular instanton
studied in Sec.~\ref{sec-closedcut2}, the boundary contribution will
be only half as large here.

For small values of $\phi_{\rm E}$, which give barely enough
inflation, this means that the absolute value of the action is largest
for the regular closed instanton of Sec.~\ref{sec-closedcut1},
followed by the gondola instanton, and the singular instanton of
Sec.~\ref{sec-closedcut2}.  For large values of $\phi_{\rm E}$, which
lead to a long period of inflation and a very flat universe, the
difference is completely negligible.  Then the gondola instanton will
be the most practical to use, since it requires no analysis of complex
variables.

\section{Summary}

We have described a number of non-singular instantons leading to open
inflating universes. They include the Coleman-De-Luccia solution, in
which a bubble of true vacuum expands inside a universe inflating in
the false vacuum. We found new solutions which contain two bubbles, or
a shell of true vacuum.

We also constructed instantons with a singularity. If the singularity
does not lie on the hypersurface of nucleation, it causes no problems
in the Lorentzian region, and can be interpreted as a small region of
Planckian density.  Such instantons can be used to describe the
quantum creation of a closed inflationary universe from space-time
foam without the need to use complex solutions.

\subsection*{Acknowledgments}

It is a pleasure to thank A. Chamblin, N. Kaloper and L.A. Kofman for
useful discussions. This work was supported in part by NSF Grant No.
PHY-9219345 and by NATO/DAAD.

\end{document}